\documentclass[twocolumn]{aastex701}
\usepackage{amsmath}
\usepackage{amssymb}
\usepackage{amsfonts}
\usepackage{graphicx}
\usepackage{multirow}

\begin{document}
\shortauthors{The SPOTLIGHT Collaboration}
\title{The SPOTLIGHT Multibeam Real-Time Transient Detection System}
\shorttitle{The SPOTLIGHT Real-time Transient Search Pipeline}

\correspondingauthor{Ujjwal Panda}
\email{ujjwalpanda97@gmail.com, upanda@ncra.tifr.res.in}
\affiliation{National Centre for Radio Astrophysics (NCRA), Pune, 411007, Maharashtra, India}

\author[0000-0002-2441-4174]{Ujjwal Panda}
\email{ujjwalpanda97@gmail.com, upanda@ncra.tifr.res.in}
\affiliation{National Centre for Radio Astrophysics (NCRA), Pune, 411007, Maharashtra, India}

\author[0000-0002-2892-8025]{Jayanta Roy}
\email{jroy@ncra.tifr.res.in}
\affiliation{National Centre for Radio Astrophysics (NCRA), Pune, 411007, Maharashtra, India}

\author[0000-0002-8550-9070]{Kshitij Bane}
\affiliation{National Centre for Radio Astrophysics (NCRA), Pune, 411007, Maharashtra, India}
\email{kshitijbane@gmail.com}

\author[0009-0005-4130-892X]{Chahat Dudeja}
\affiliation{National Centre for Radio Astrophysics (NCRA), Pune, 411007, Maharashtra, India}
\email{dudejachahat11@gmail.com}

\author[0000-0003-3747-9847]{Sridhar Gajendran}
\affiliation{National Centre for Radio Astrophysics (NCRA), Pune, 411007, Maharashtra, India}
\email{sridhar.gajendran@gmail.com}

\author[0009-0005-9293-0655]{Param Joshi}
\affiliation{National Centre for Radio Astrophysics (NCRA), Pune, 411007, Maharashtra, India}
\affiliation{Department of Physics and Astronomy, National Institute of Technology, Rourkela, 769008, India}
\email{paramjoshi1812@gmail.com}

\author[0000-0002-6631-1077]{Sanjay Kudale}
\affiliation{National Centre for Radio Astrophysics (NCRA), Pune, 411007, Maharashtra, India}
\email{kudale.sanjay@gmail.com}

\author[0009-0007-8409-4233]{Arpan Pal}
\affiliation{National Centre for Radio Astrophysics (NCRA), Pune, 411007, Maharashtra, India}
\email{apal@ncra.tifr.res.in}

\author[0009-0002-2515-2425]{Raghav Wani}
\affiliation{National Centre for Radio Astrophysics (NCRA), Pune, 411007, Maharashtra, India}
\affiliation{Indian Institute of Science Education and Research (IISER), Pune, 411008, Maharashtra, India}
\email{raghav.wani@gmail.com, wani.raghav@students.iiserpune.ac.in}

\author[0000-0003-2797-0595]{Karel Adamek}
\affiliation{Department of Physics, Silesian University in Opava, Opava, 74601, Czech Republic}
\email{karel.adamek@gmail.com}

\author[0000-0003-1756-3064]{Wes Armour}
\affiliation{Oxford e-Research Centre (OeRC), University of Oxford, Oxford-OX13PJ, United Kingdom}
\email{wes.armour@oerc.ox.ac.uk}

\author[0009-0003-8718-1560]{Kenil Ajudiya}
\affiliation{International Centre for Radio Astronomy Research (ICRAR), Curtin University, Bentley, WA 6102, Australia}
\email{kra635900@gmail.com}

\author[0009-0005-4130-892X]{Jayaram Chengalur}
\affiliation{National Centre for Radio Astrophysics (NCRA), Pune, 411007, Maharashtra, India}
\email{chengalur@ncra.tifr.res.in}

\author[0009-0006-7995-5871]{Jyotirmoy Das}
\affiliation{National Centre for Radio Astrophysics (NCRA), Pune, 411007, Maharashtra, India}
\email{tataidas5392@gmail.com}

\author[0009-0001-4178-3879]{Nishant Pradeep Deo}
\affiliation{National Centre for Radio Astrophysics (NCRA), Pune, 411007, Maharashtra, India}
\affiliation{Indian Institute of Science Education and Research (IISER), Kolkata, Mohanpur-741246, West Bengal, India}
\email{npd21ms206@iiserkol.ac.in}

\author[0009-0000-9152-3832]{Deepak Bhong}
\affiliation{National Centre for Radio Astrophysics (NCRA), Pune, 411007, Maharashtra, India}
\email{deepak@gmrt.ncra.tifr.res.in}

\author[0000-0000-0000-0000]{Shelton Gnanaraj}
\affiliation{National Centre for Radio Astrophysics (NCRA), Pune, 411007, Maharashtra, India}
\email{shelton@gmrt.ncra.tifr.res.in}

\author[0009-0007-4654-4539]{Santaji N. Katore}
\affiliation{National Centre for Radio Astrophysics (NCRA), Pune, 411007, Maharashtra, India}
\email{snk@gmrt.ncra.tifr.res.in}

\author[0009-0008-1233-6915]{Mekhala Muley}
\affiliation{National Centre for Radio Astrophysics (NCRA), Pune, 411007, Maharashtra, India}
\email{mekhala@gmrt.ncra.tifr.res.in}

\author[0000-0002-7551-5215]{Harshavardhan Reddy}
\affiliation{National Centre for Radio Astrophysics (NCRA), Pune, 411007, Maharashtra, India}
\email{reddysh@gmrt.ncra.tifr.res.in}

\begin{abstract}
Fast Radio Bursts (FRBs) are among the most enigmatic transient phenomena in the Universe. In order to unravel the mystery behind these extremely short-lived events, one requires instruments that possess the ability to detect, localise, and capture these events high spectro-temporal-polarimetric resolution, as they search for these events over large fields-of-view in real-time. The SPOTLIGHT project is one such backend, leveraging the high sensitivity, wide frequency coverage, and excellent localisation capabilities of the upgraded Giant Metrewave Radio Telescope (uGMRT) to conduct a commensal search for FRBs and other radio transients. This search is done using a dedicated high-performance computing facility, comprised of 90 NVIDIA A100 GPUs and 60 compute servers. Here we present the design, implementation, and performance of SPOTLIGHT's real-time transient search pipeline, a GPU-accelerated system capable of processing up to 2000 post-correlation beams in real time. The pipeline combines \texttt{AstroAccelerate}-powered brute-force dedispersion and single-pulse search, with a multi-stage and robust candidate optimisation framework, as well as a triggering system for automatic capture of high-time-resolution visibility and baseband data. To ensure continuous validation of pipeline performance, we have also developed a real-time signal injection framework capable of injecting synthetic bursts directly into SPOTLIGHT's beamformed data stream. The system operates commensally with routine uGMRT observations, processing data streams in real-time while maintaining high sensitivity to millisecond-duration transients across dispersion measures extending up to 2000 pc cm$^{-3}$. During its initial deployment in uGMRT Cycle 49 and Cycle 50, the pipeline detected 2870 bursts from 42 known astrophysical sources, and demonstrated sensitivity consistent with the predicted survey threshold of $\sim$0.2 Jy ms. SPOTLIGHT's real-time transient search pipeline establishes a scalable framework for wide-field, low-frequency transient discovery and localisation, providing a key technological foundation for next-generation radio transient surveys.
\end{abstract}
\section{Introduction}

In time domain astronomy, fast radio bursts (FRBs) stand out as one of the most mysterious sources. Radiating on timescales of a few milliseconds or shorter, their emission is bright, coherent, wideband (from 110 MHz \cite{pleunis_lofar_2021} to 8 GHz \cite{gajjar_highest_2018}), and highly polarised. The sources themselves are compact, as well as extragalactic. The former was inferred from the timescale and coherency of their emission, while the latter was inferred from their large dispersion measures. Both properties were confirmed through VLBI-based imaging and localisation of these sources to their host galaxies \citep{chatterjee_direct_2017, tendulkar_host_2017, bannister_single_2019, xu_fast_2022}. Since their serendipitous discovery in 2007 \cite{lorimer_bright_2007}, their number has grown drastically, especially after the inception of the CHIME/FRB project using the Canadian Hydrogen Intensity Mapping Experiment (CHIME) telescope \citep{chimefrb_collaboration_chime_2018}. Originally meant for mapping 21-cm emission across the radio sky with its wide field-of-view, this very feature coincidentally made it into the most prolific FRB discovery instrument, single-handedly contributing 3641 FRBs, or $\sim 72.7\%$ of the total fraction, which now total up to 5006 (estimated by combining the catalogue of FRBs maintained at the Transient Name Server (TNS), \url{https://www.wis-tns.org}, with the catalogue maintained at Blinkverse, \url{https://blinkverse.zero2x.org}). However, the number of FRBs localised to their host galaxies has been low (119, or only $\sim 2.38\%$ of FRBs). It is to be noted that while this precision has since been improved through the dumping of baseband data \citep{michilli_analysis_2021}, and subsequently through the construction of outrigger stations \citep{lanman_chimefrb_2024, khairy_green_2024}, a majority (3615 out of 3641, or 99.29\%) of the detections in CHIME's catalogue remain \textit{poorly localised}; that is, with a localisation precision $\lesssim 5$ arcseconds that is required to associate an FRB unambiguously to a host galaxy with a redshift $z \gtrsim 0.1$ \citep{eftekhari_associating_2017}. This prompted commensal surveys from several interferometers despite their small fields-of-view, such as CRAFT \citep{macquart_commensal_2010} and CRACO \citep{wang_craft_2025} by ASKAP, MeerTRAP \citep{sanidas_meertrap_2017} by MeerKAT, and realfast \cite{law_realfast_2018} by the VLA. Dedicated interferometers were commissioned as well, such as DSA-110 \citep{law_deep_2023}, and BURSTT \citep{lin_burstt_2022}. This has led to an increase in localisations and successful associations of both repeating \textit{and} one-off FRBs to their host galaxies. Pinning down an FRB to its host provides a deeper look into its local environment \citep{tendulkar_60_2021}, possible progenitors models \citep{metzger_millisecond_2017, metzger_fast_2019, margalit_constraints_2020}, and enables its use as a probe of both cosmology \citep{macquart_census_2020, li_cosmology-insensitive_2020}, and the intergalactic medium (IGM) \citep{simha_disentangling_2020, sammons_two-screen_2023}.

For the reasons outlined above, the SPOTLIGHT project (\textbf{S}urvey for s\textbf{P}oradic radi\textbf{O} burs\textbf{T}s via a commensa\textbf{L} mult\textbf{I}-beam \textbf{G}PU-powered \textbf{H}PC at the GMR\textbf{T}) \footnote{\url{https://spotlight.ncra.tifr.res.in}} was commissioned (Roy et al. (in prep)). It aims to leverage the upgraded GMRT's high sensitivity, large bandwidths, unique low-frequency spectral windows, and localisation capabilities to carry out a commensal survey of FRBs and other transients. An overview of the SPOTLIGHT project will be provided in Roy et al. (in prep). In order to carry out a survey commensally with regular GMRT observations (over 300$-$1460 MHz frequency range), a dedicated backend was built, funded under the Government of India's National Supercomputing Mission (NSM). Known as the SPOTLIGHT system, it consists of 60 of C-DAC's indigenously developed Rudra servers, each of which houses a 768/128 GB RAM, a Intel(R) Xeon(R) Gold 6240R CPU, with 24 physical and 48 logical cores, and 2 NVIDIA A100 GPUs, each with 80 GB of GPU memory. Overall, the SPOTLIGHT cluster has 90 NVIDIA A100 GPUs, $\sim 40$ TB of CPU memory, $\sim 7$ TB of GPU memory, and 2 PB of on-disk storage. It takes in data from GMRT's digitisers at a rate of 25 GB/s, via 32 10 GB fibre links. This data is then processed by SPOTLIGHT's correlator and beamformer (Reddy et al. (in prep)) to form up to 2000 post-correlation (PC) beams \cite{roy_post-correlation_2018}, through the phased addition of visibilities at a time resolution of 1.3 milliseconds. These beams are then optimally tiled, balancing the field-of-view (FoV) with sensitivity. An incoherent beam is also formed using the autocorrelation terms rejected when forming the post-correlation beams, allowing for a much wider FoV while sacrificing sensitivity. The system holds $\left(1.31072 \times 10^{-3}\right) \times 800 \times 32 \times 12 = 402.653184$ seconds of data in its transient, visibility, and baseband buffers simultaneously. The beamformed and visibility data are sampled every 1.31072 milliseconds, while the baseband data is sampled at the Nyquist rate (2.5 nanoseconds for a bandwidth of 200 MHz, and 1.25 nanoseconds for a bandwidth of 400 MHz). Once a trigger is generated by the real-time transient search pipeline, its corresponding data is dumped from all buffers to disk. The 1.3 ms time resolution visibilities are then used to image each event, via our real-time imaging pipeline (Pal et al. (in prep)).

An overview of SPOTLIGHT's multibeam and incoherent beam real-time transient search pipelines is presented in \S\ref{sec:overview}, and a module-by-module description is given in \S\ref{sec:modules}. The performance of the pipeline is discussed in \S\ref{sec:performance}, and results obtained from the pipeline are presented in \S\ref{sec:results}. Finally, we summarise in \S\ref{sec:summary}.
\section{Pipeline Overview}\label{sec:overview}

\begin{figure*}
    \centering
    \includegraphics[width=0.85\textwidth]{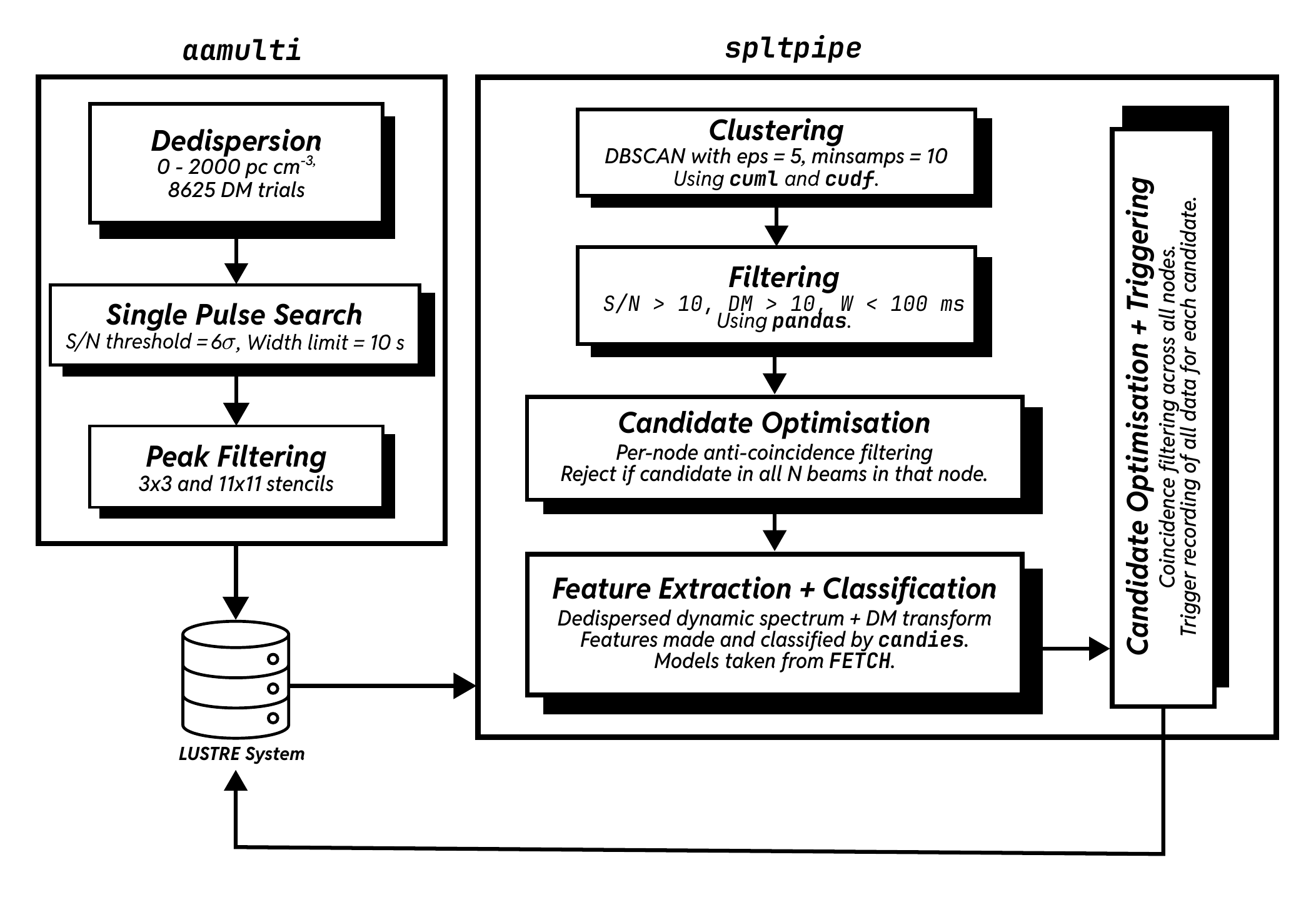}
    \caption{A flowchart illustrating how SPOTLIGHT's real-time transient search pipeline is structured. The pipeline is divided into two modules: \texttt{aamulti} and \texttt{spltpipe}, in order to maximise GPU utilisation and throughput. All results are written to the LUSTRE filesystem, mounted across all hosts. A detailed description of individual modules can be found in the text.}
    \label{fig:flowchart}
\end{figure*}

SPOTLIGHT's real-time transient search pipeline utilises 32 NVIDIA A100 GPUs, in conjunction with 768 CPU cores, across 16 Rudra servers. The processing framework is divided into two asynchronous pipelines, \texttt{aamulti} and \texttt{spltpipe}, designed to maximise hardware utilisation by keeping both GPUs on each compute node continuously occupied. This architecture can sustain a computational throughput of ($\sim$ 252 TFLOPS) while processing 2000 beams in real time, corresponding to nearly 80\% of the theoretical peak single-precision performance of 32 NVIDIA A100 GPUs. The former half is described in \S\ref{sec:aamulti}, while the latter is described in \S\ref{sec:spltpipe}. A bird's-eye view of the entire pipeline can be seen in Fig. \ref{fig:flowchart}. Currently, the SPOTLIGHT system is deployed with 640 post-correlation (PC) beams, at 1.31072 milliseconds time resolution, and 4096 frequency channels. An incoherent array (IA) beam, with the same time and frequency resolution, is formed in parallel, and is searched independently for transient events using \texttt{aa.single} and \texttt{splt.single}, single beam analogues of the same pipeline used for the PC beams; this pipeline is described in more detail in \S\ref{sec:iabeam}.

\begin{figure*}
    \centering
    \includegraphics[width=0.85\textwidth]{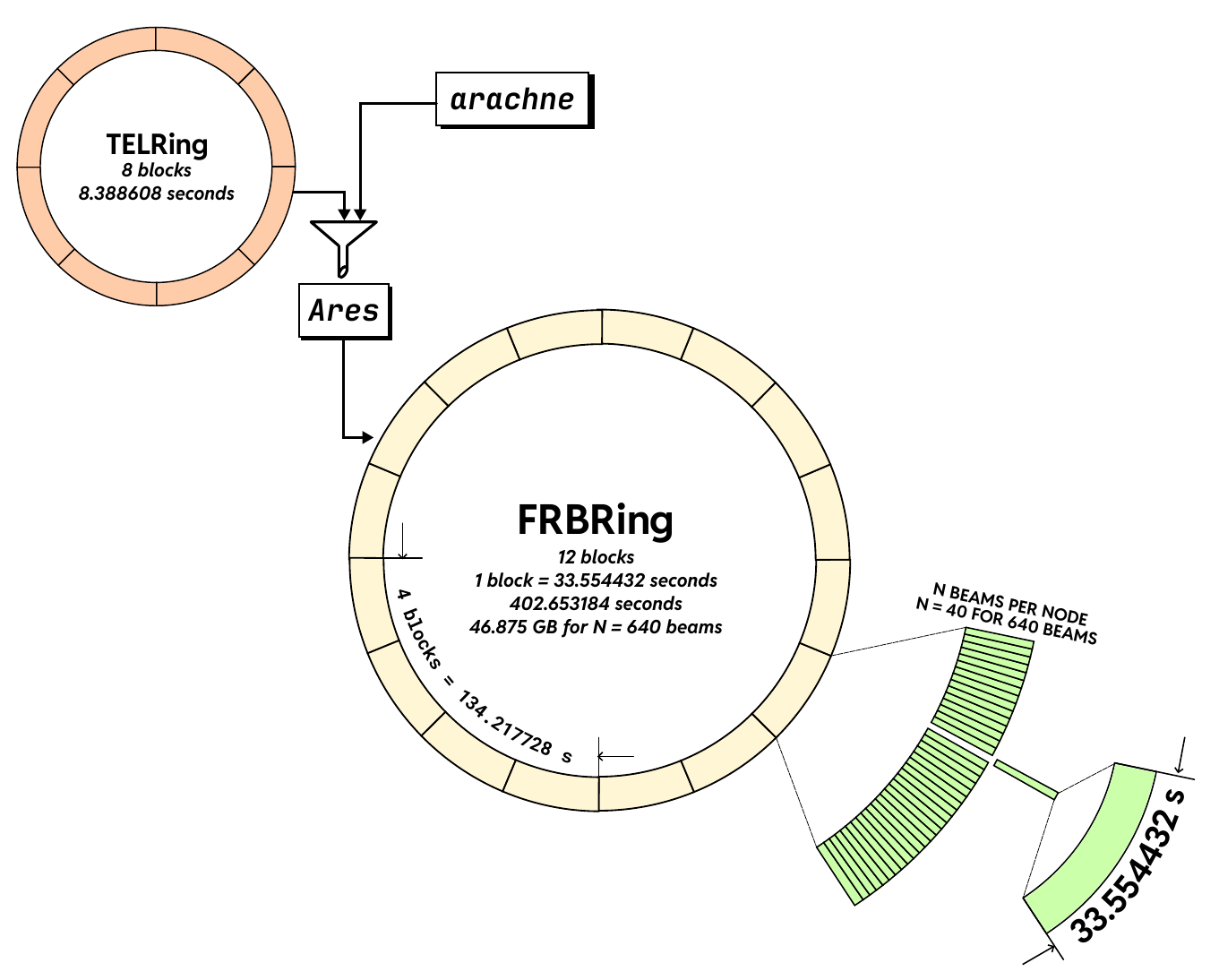}
    \caption{SPOTLIGHT manages beamformed data using two shared-memory-based ring buffers: \texttt{TELRing} and \texttt{FRBRing}. The former is where data is written to by SPOTLIGHT's correlator/beamformer, after which data is appropriately processed and transferred to the latter by \texttt{Ares}. Simulated FRB-like signals can also be injected during this transfer, using \texttt{arachne}. The pipeline reads data from \texttt{FRBRing}, and processes 4 blocks at a time. Each block is time-spliced to contain the data from 40 beams, across 16 hosts, for a total of 640 beams.}
    \label{fig:ringbuffers}
\end{figure*}

Beamformed data obtained from SPOTLIGHT's correlator/beamformer (Reddy et al. in preparation) is first written to a small ring buffer, known internally as \texttt{TELRing}. It is divided into 8 blocks; each block is $800 \times 1.31072 \times 10^{-3} = 1.048576$ seconds long, making the entire buffer only 8.388608 seconds long. Data is read from this ring buffer one block at a time, appropriately scaled, radio frequency interference (RFI) mitigated, flipped (so that the highest frequency is first; required for certain observation bands at the GMRT which start from the lowest frequency channel instead), injected with simulated events (if specified; see \S\ref{sec:arachne} for more details), and written to a larger ring buffer, known as \texttt{FRBRing}. All of this is handled by a dedicated program called \texttt{Ares}\footnote{\url{https://github.com/nsmspotlight/Ares}}. The RFI mitigation methodology is inspired from \texttt{PulsarX}'s \texttt{filtool} program \citep{men_pulsarx_2023}, and will be described in more detail in Wani et al. (in prep). Both \texttt{aamulti} and \texttt{spltpipe} read data from this \texttt{FRBRing} buffer. This buffer, created by \texttt{Ares}, is divided into 12 blocks, and each one of them is $1.048576 \times 32 = 33.554432$ seconds long; that is, the same duration as 32 blocks of \texttt{TELRing}. Each block holds the time-spliced data from $N$ beams, where $N$ is the total number of beams divided by the number of hosts; currently $N = 640 / 16 = 40$. This implies that \texttt{FRBRing} holds $33.554432 \times 12 = 402.653184$ seconds, or 46.875 GB of data, for 640 beams in memory at any given time. The same duration of data is held in analogous ring buffers for high-resolution visibility data (sampled at 1.31072 milliseconds time resolution), and baseband data (sampled at the Nyquist resolution). Holding a large duration of data in memory is necessary, since it allows us to dump beamformed, visibility, and baseband data for each event once it is detected, while taking any pipeline delays into account. More details on how this data dumping is triggered is given in \S \ref{sec:triggering}. A diagram illustrating the structure of \texttt{TELRing} and \texttt{FRBRing}, and how data is transferred between them, can be seen in Fig. \ref{fig:ringbuffers}. The ring buffers for incoherent beam data follow the same design, except that there is no time-splicing required, since there is only a single beam in this case. All of these ring buffers can be accessed in both C++ and Python via the \texttt{shazam}\footnote{\url{https://github.com/astrogewgaw/shazam}} library.

\section{Pipeline Modules}\label{sec:modules}

\subsection{\texttt{aamulti}}\label{sec:aamulti}

The first half, \texttt{aamulti}, is powered by the \texttt{AstroAccelerate}\footnote{\url{https://github.com/AstroAccelerateOrg/astro-accelerate}} library. The pipeline reads $4 \text{ blocks} = 134.217728 \text{ seconds}$ at a time, and searches for single pulses. To accommodate the maximum dispersion delay corresponding to a DM of 2000 pc cm$^{-3}$ across Band 3 (300$-$500 MHz), the pipeline employs an overlap-read strategy of one data block ($= 33.554432$ seconds) between consecutive processing segments, ensuring highly dispersed signals that cross block boundaries can be recovered completely. The metadata is read from a separate shared memory buffer, which is updated every 1.048576 seconds. The search done by \texttt{aamulti} has two stages: dedispersion and single pulse search. The former is described in more detail in \S\ref{sec:dedispersion}, and the latter in \S\ref{sec:sps}.

\subsubsection{Dedispersion}\label{sec:dedispersion}

As a radio signal passes through the interstellar medium (ISM) along the line-of-sight between the Earth and the source, different frequencies travel at different speeds due to their interactions with the ionised plasma that comprises the ISM; this effect is called \textit{dispersion}. Before we can search for such signals, this effect needs to be undone. Assuming that the signal travels through cold plasma, we can calculate the amount of delay introduced for a particular frequency, with respect to some reference frequency, as:

\begin{align}\label{eq:dmdelay}
\Delta \tau 
  &= \left(\frac{e^{2}}{2\pi m_{e} c}\right)
  \times \mathrm{DM}
  \times \left(
    \frac{1}{f^{2}}
    - \frac{1}{f_{\mathrm{high}}^{2}}
  \right) \\
  &= \mathcal{D}
  \times \mathrm{DM}
  \times \left(
    \frac{1}{f^{2}}
    - \frac{1}{f_{\mathrm{high}}^{2}}
  \right),
\end{align}
    
where $f$ is the frequency for which the delay is being calculated, $f_{\mathrm{high}}$ is the highest frequency in the observation band, which is taken as the reference frequency, $\mathcal{D} = \frac{e^{2}}{2\pi m_{e} c} = 4.1488064239(11) \times 10^{3} \text{ MHz}^{2} \text{ pc}^{-1} \text{ cm}^{3}$ s is the dispersion constant \citep{kulkarni_dispersion_2020}, and DM is the \textit{dispersion measure}. The dispersion measure quantifies the total amount of plasma encountered by the signal, and is defined as the integrated column density of electrons along the line-of-sight; that is:

\begin{equation}
    \mathrm{DM} = \int n_{e} \, dl,
\end{equation}

\begin{figure*}
    \centering
    \includegraphics[width=\textwidth]{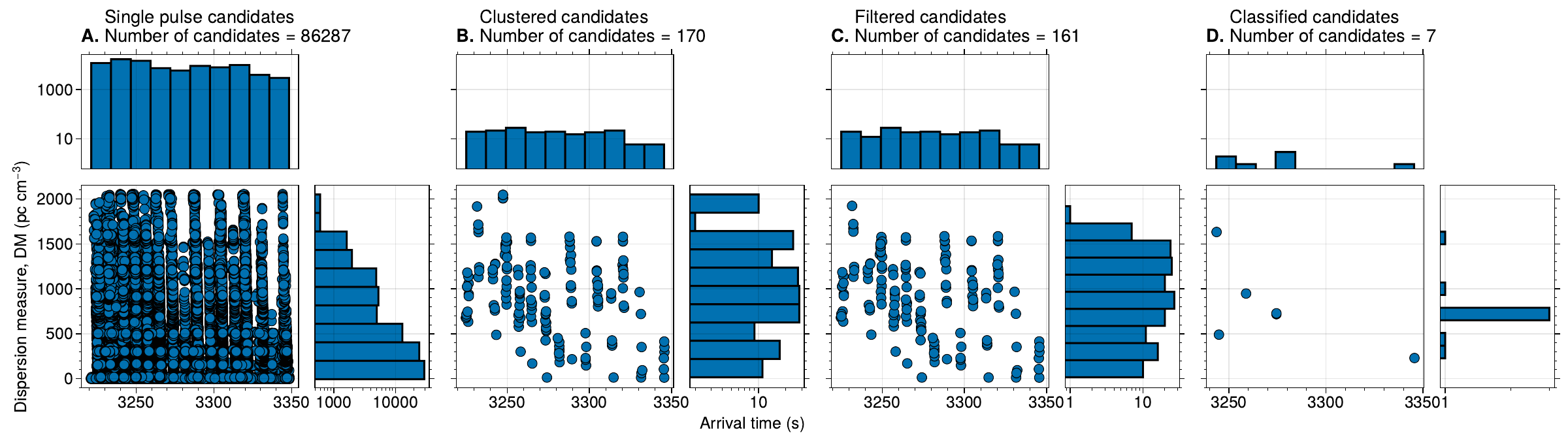}
    \caption{Scatter plots of the dispersion measure (DM) versus the arrival time for events detected in a single beam from a single time block of 134.217728 seconds, as seen at each stage of SPOTLIGHT's transient search; from left to right, these stages are A) single pulse search, B) clustering, C) filtering, and D) classification. These plots illustrate how spurious candidates are eliminated at each stage, starting from 86,287 candidates at the first, single pulse search stage, to only 7 candidates at the final, classification stage.}
    \label{fig:pipeline_stages}
\end{figure*}

where $n_{e}$ is the column density of electrons, and $dl$ is a line element along the line-of-sight. The process of removing or mitigating these effects from the signal before processing is called dedispersion. However, since the DM along the line-of-sight is not known a priori in a blind search, dedispersion is carried out for a large number of trial DM values. Note that dedispersing a signal at an incorrect DM smears it, and with enough smearing it can become undetectable. Thus, the trial DM values are chosen so that this smearing is kept at a minimum, while also keeping the number of trials to a minimum. This results in a dedispersion plan, which divides the entire DM range one wishes to search over into one or more sub-ranges; within each sub-range, trial DM values are spaced so as to minimise smearing, which is estimated based on the parameters of the observation setup; these are the sampling time, the frequency resolution, and the observation bandwidth. Such a plan has been constructed for the SPOTLIGHT survey, with a maximum DM limit of 2000 pc cm$^{-3}$ for GMRT's Band 3 (300 $-$ 500 MHz), Band 4 (550 $-$ 750 MHz), and Band 5 (1060 $-$ 1460 MHz), covered with 8625 DM trials.
    
One can dedisperse a signal via one of two approaches: coherent or incoherent dedispersion. In the coherent dedispersion, one corrects for the dispersion in the Fourier domain, where it is a phase-only filter, and thus one can dedisperse the signal by multiplying its spectrum with the filter's inverse. However, while this approach boasts both high accuracy and sensitivity, it is also quite compute-intensive, due to the large FFTs that need to be evaluated for each DM trial. Hence, blind surveys typically opt for incoherent dedispersion, where the effect is corrected by shifting each channel in channelised intensity data (also known as a \textit{dynamic spectrum}) by the appropriate time delay, calculated using Eq. \ref{eq:dmdelay}. Multiple approaches exist which approximate this correction in favour of speed, such as subband dedispersion \citep{}, tree dedispersion \citep{}, or the fast DM transform \citep{zackay_accurate_2017}. However, these approximations trade-off sensitivity, accuracy, or flexibility for speed. Thus, in \texttt{aamulti} we instead opt for the GPU-based implementations of brute-force dedispersion that come packaged with the \texttt{AstroAccelerate} library. The library provides two implementations: one uses a GPU's cache memory, and another that utilises a GPU's shared memory \citep{novotny_accelerating_2023}. The latter implementation leverages the faster memory access and data locality of a GPU's shared memory; and since dedispersion is ultimately a \textit{memory-bound} algorithm, rather than a \textit{compute-bound} one, these optimisations are crucial to accelerating dedispersion on multi-core architectures. The former version is a fail-safe, used when the latter implementation cannot be used; this usually happens when the data cannot fit into the limited size of a GPU's shared memory. \texttt{aamulti}'s dedispersion plans in each band are carefully designed to ensure that the fail-safe system is never invoked. The output from either of the two implementations is a two-dimensional array, with each row representing a time series at a particular trial DM value. This array is known as a \textit{DM transform}.

\subsubsection{Single pulse search}\label{sec:sps}

After creating the DM transform, \texttt{aamulti} searches each row for single pulses. The methodology used for this search was first given in \cite{cordes_searches_2003}, and is based on the matched filtering theorem. The theorem guarantees that if the input to a filter, $x(t)$, \textit{matches} its impulse response, $h(t)$\footnote{The impulse response is how a filter responds to a delta function, $\delta(t)$. Since any signal can be reconstructed as a series of delta functions, the impulse response describes the response of a filter to \textit{any} function; in fact, the output of the filter, for any input $x(t)$, is $y(t) = x(t) \otimes h(t)$.}, then the signal-to-noise ratio (S/N) of the filter's output, $y(t)$, will be maximum. For a linear and time-invariant filter, this output is merely the convolution of the input signal with the impulse response; that is, $y(t) = x(t) \otimes h(t)$. If we think of $h(t)$ as a \textit{template}, then we use this method to search for single pulses, since wherever we get a pulse in the input signal that matches our template, our output's S/N will rise, and we can then pick all pulses in the output above a certain S/N threshold as our detections. In order to optimise this method further, one uses a boxcar as a template, since the convolution of any signal with a boxcar is the cumulative sum of the input signal across the boxcar width. Since the width of the signal is not known a priori, the search is carried using multiple boxcar filters, with different trial widths. \texttt{aamulti} uses \texttt{AstroAccelerate}'s GPU-optimised single pulse search algorithm, as described in \cite{adamek_single-pulse_2020}. The single pulse search is carried out up to a maximum width of 0.5 seconds, and all candidates above a threshold of 6$\sigma$ are taken to the next stage. At this point, the number of candidates per beam can range from $10^{3}$ to $10^{6}$.
    
After searching for single pulses, the \texttt{AstroAccelerate} library can employ another stage, that of peak filtering, which leads to an order-of-magnitude reduction in the number of candidates. Peaks are filtered out by sliding a window across the candidates in the DM-time plane. From each window, the peak with the highest S/N is picked, and all others are rejected. Two different stencils are used: a $3 \times 3$ window, and an $11 \times 11$ window. The number of candidates per beam at this stage can range from $10^{2}$ to $10^{6}$ per beam per block.

\begin{figure*}
    \centering
    \includegraphics[width=\textwidth]{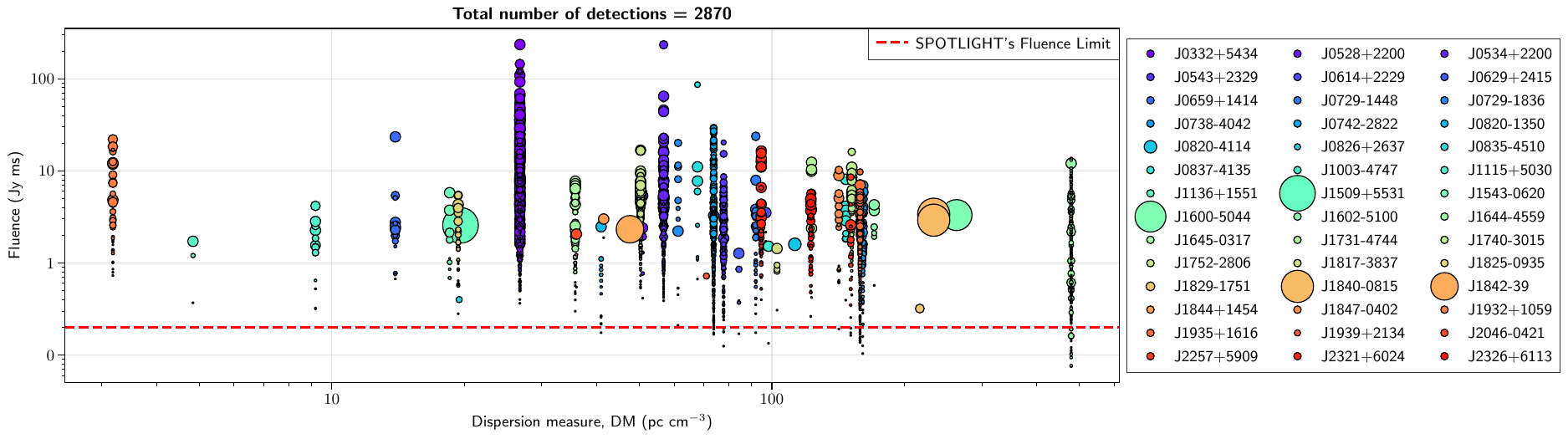}
    \caption{A census of detections obtained from SPOTLIGHT's real-time transient search pipeline via commensal observations in GMRT's Cycle 49 and Cycle 50. A total of 2870 detections were made from 42 known sources; each colour in the plot represents a particular source. The fluence of each detected bursts is plotted, versus its dispersion measure (DM). Each point's size is scaled by its detected width. The red dashed line is the predicted fluence limit of 0.2 Jy ms (for a burst with an S/N 0f 10) for the SPOTLIGHT survey.}
    \label{fig:census}
\end{figure*}

\subsection{\texttt{spltpipe}}\label{sec:spltpipe}

The second half, \texttt{spltpipe}, responds as soon as \texttt{aamulti} is done processing. Despite employing a peak-filtering stage in \texttt{aamulti}, the number of candidates obtained can range from a thousand to around a million per beam, depending on the RFI environment. A majority of these candidates are often spurious. Thus, \texttt{spltpipe} takes the output obtained from \texttt{aamulti}, and removes spurious candidates in several stages. An overview of how this removal happens at each stage can be seen in Fig. \ref{fig:pipeline_stages}.

\subsubsection{Candidate optimisation}\label{sec:optimisation}

Candidates from each beam are clustered in the DM-time plane using the DBSCAN (Density-Based Spatial Clustering of Applications with Noise, \cite{ester_density-based_1996}) algorithm. A cluster is only considered if it has 5 points or more in the DM-time plane. The neighbourhood radius parameter, ($\epsilon$), is set to 0.5 based on the expected distribution of astrophysical and RFI detections in the DM–time plane, taking into account the sensitivity and frequency coverage of the GMRT observations. Out of each cluster, the peak with the highest S/N is picked, and all other peaks are rejected. The number of candidates at this stage can range from $10^{2}$ - $10^{3}$ per beam per block. After clustering, the candidates from all beams in a particular host are combined together into one list, and are then passed through several thresholds, based on DM, width, and S/N. Currently, \texttt{spltpipe} rejects all candidates below an S/N of 10, a DM of 10 pc cm$^{-3}$, and higher than a width of 100 ms. These thresholds are configurable, even during a running observation. Spurious candidates can be reduced further by connecting candidates \textit{across beams}. We would like to describe this approach as (anti-)coincidence filtering, wherein associating candidates detected at the same arrival time across neighbouring beams would be deemed as \textit{coincidence}, whereas rejecting candidates detected at the same arrival time, but across vastly different and/or scattered beams would be deemed \textit{anti-coincidence}. The latter is facilitated on a per-node basis by distributing beams randomly across the nodes, so that beams that end up on the same node have vastly different sky positions. Thus, a candidate that is present in all, or a majority, of the beams is definitely RFI, and can be removed. The number of candidates at this stage can range from 10 to $10^{2}$ per beam per block.
    
\subsubsection{Classification}\label{sec:classification}

For each candidate that survives the candidate optimisation procedure laid out in \S\ref{sec:optimisation}, the beamformed data corresponding to it is extracted from the ring buffers described in \S\ref{sec:overview}; this is taken care of via \texttt{shazam}. It extracts a number of samples equal to the maximum dispersive delay across the observation band, plus the number of samples corresponding to the width of the candidate, both before and after the candidate's arrival time. This data is then given to \texttt{candies}\footnote{Source code: \url{https://github.com/astrogewgaw/candies}. The package is also available on the Python Packing Index (PyPI) here: \url{https://pypi.org/project/candies}.}, which extracts two features from this data: the dedispersed dynamic spectrum, and the DM transform. \texttt{candies} replaces \texttt{your}~\footnote{\url{https://github.com/thepetabyteproject/your}}, which is what was previously used to generate these features. The former is created at the candidate's detected DM, while the latter is created at a narrow DM range centred around that DM. In \texttt{your}, this range spanned 0 pc cm$^{-3}$ to twice the candidate's DM. In \texttt{candies} this range is calculated by estimating the change in DM required in order to decrease S/N by a certain factor; this change is estimated using the following equation:

\begin{equation}\label{eq:deltadm}
    \delta\mathrm{DM}
    \approx \frac{\sqrt{\pi}}{1382}
    W_{\mathrm{s}}
    \left(\frac{S}{S(\delta DM)}\right)
    \frac{\nu_{\mathrm{MHz}}^{3}}{\Delta\nu_{\mathrm{MHz}}},
\end{equation}

where $\Delta\nu_{\mathrm{MHz}}$ is the observation bandwidth in MHz, $\nu_{\mathrm{GHz}}$ is the central observational frequency in GHz, $W_{\mathrm{ms}}$ is the width of the candidate in milliseconds, and $\frac{S(\delta DM)}{S}$ is the ratio of the S/N of the pulse dedispersed at the wrong trial DM, versus the peak S/N. By default, this ratio is set to $0.1$ in \texttt{candies}, corresponding to a $90\%$ loss in S/N with respect to the peak S/N of the burst. A detailed derivation of Eq. \ref{eq:deltadm} can be found in Appendix \ref{appendix:A}. 

\begin{figure*}
    \centering
    \includegraphics[width=0.85\textwidth]{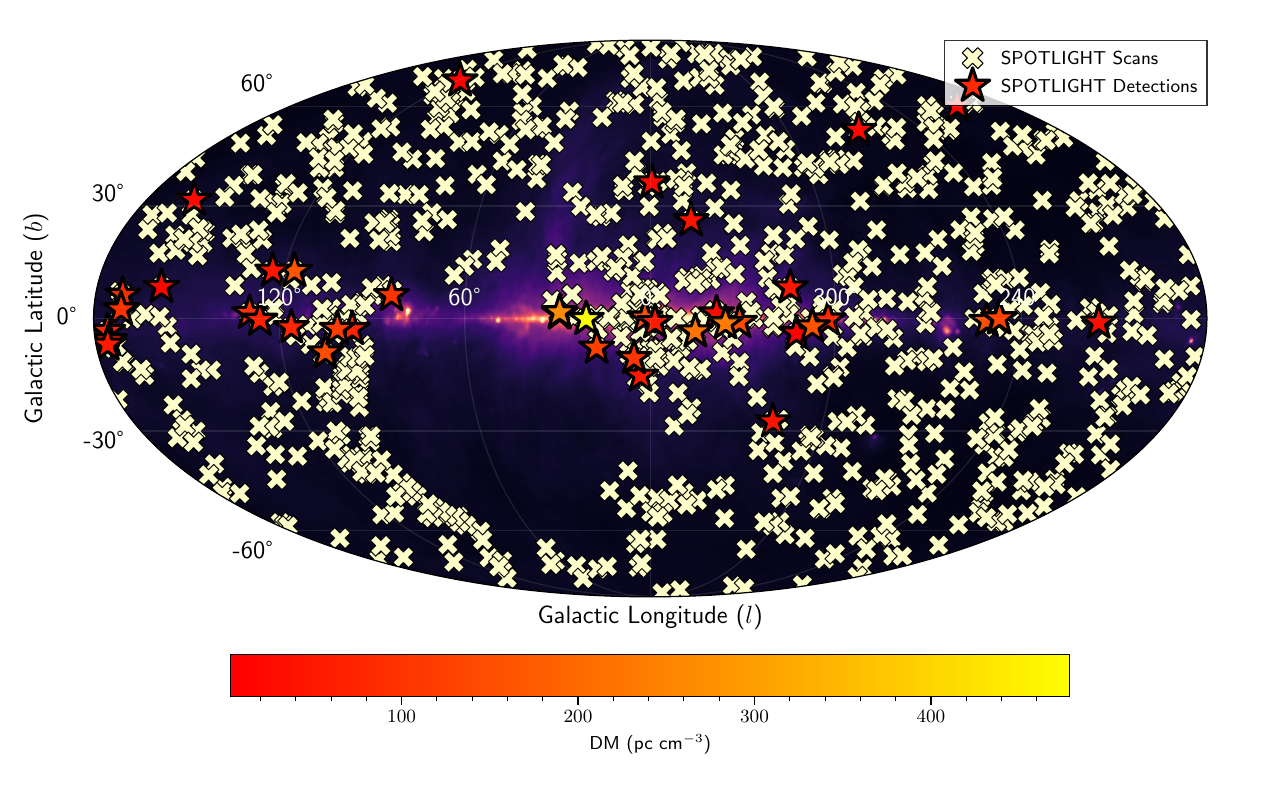}
    \caption{A galactic map of the sources detected obtained from SPOTLIGHT's real-time transient search pipeline via commensal observations in GMRT's Cycle 49 and Cycle 50. A total of 39 unique sources were detected. The colour represents the dispersion measure (DM) of each source.}
    \label{fig:galactic_map}
\end{figure*}

\begin{figure*}
    \centering
    \includegraphics[width=0.85\textwidth]{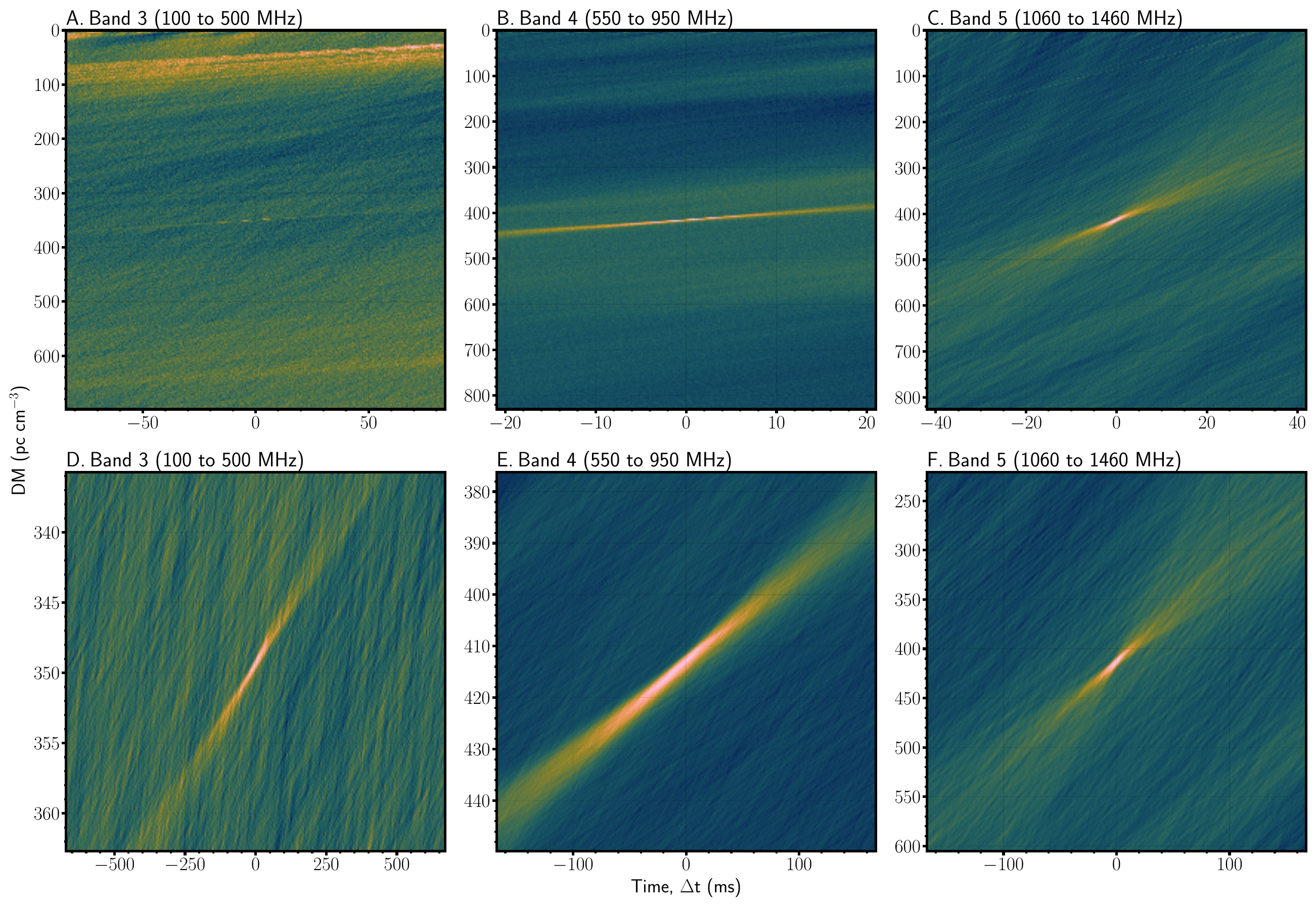}
    \caption{Variation of the DM transform (DMT) over different frequency bands at the GMRT, for actual bursts detected from FRB~20180916B and FRB~20201124A. From left to right, we have Band 3 (300 to 500 MHz), Band 4 (550 to 750 MHz), and Band 5 (1060 to 1460 MHz). For real signals, we expect a bow-tie pattern in the DMT, which becomes more and more difficult to observe at lower frequencies. Plots in the first row were created using \texttt{your}'s \texttt{your\_candmaker.py} script, while plots in the second row were created via \texttt{candies} for the same exact bursts.}
    \label{fig:your_vs_candies_zoom}
\end{figure*}

\begin{figure*}
    \centering
    \includegraphics[width=0.85\textwidth]{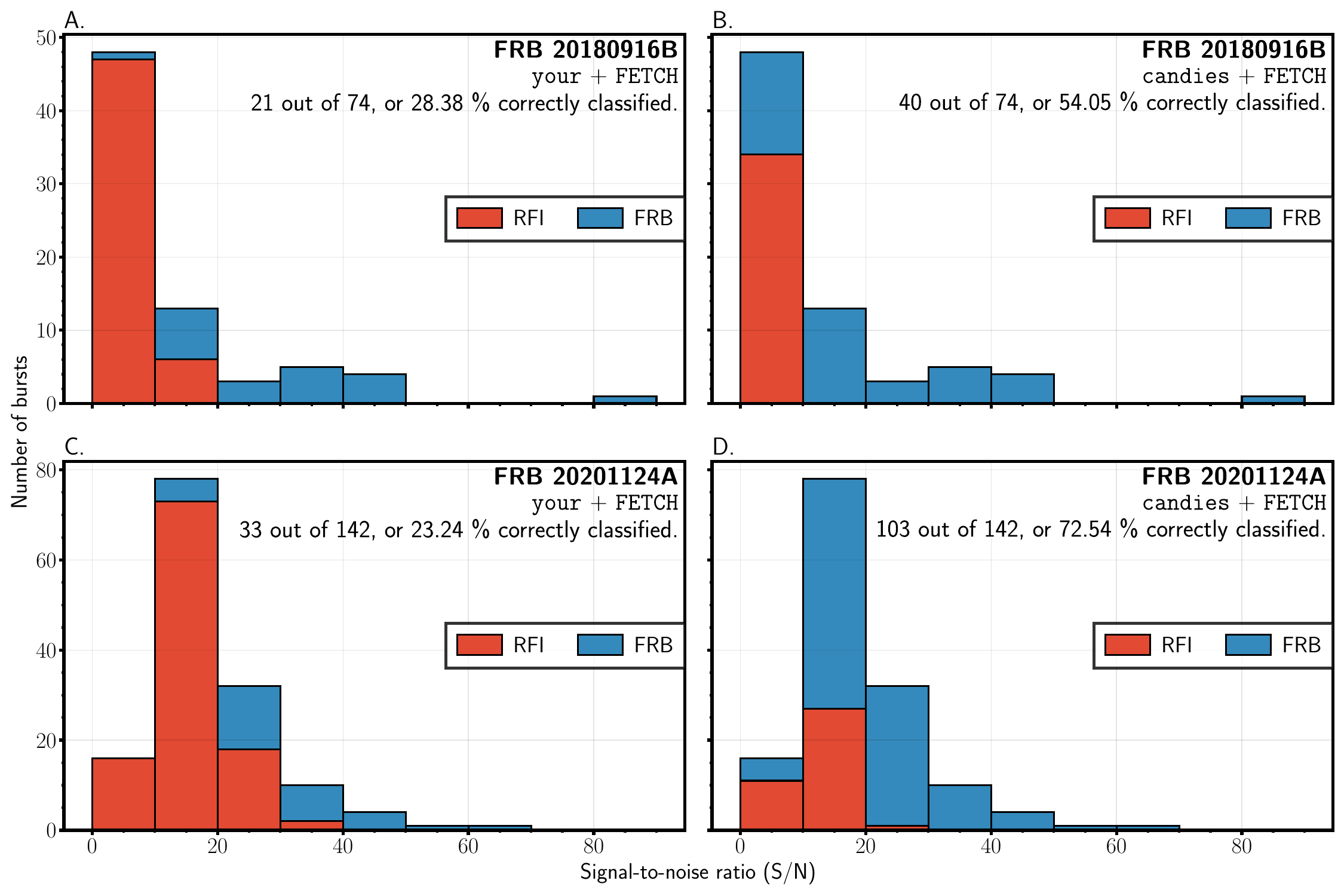}
    \caption{Histograms of S/N for 74 bursts detected from FRB~20180916, and 142 bursts detected from FRB~20201124, using the GMRT. These bursts were passed through both \texttt{your} + \texttt{FETCH} (\textit{left}) and \texttt{candies} + \texttt{FETCH} (\textit{right}). The bursts wrongly labelled as RFI are in red, while the bursts correctly labelled as FRBs are in blue. From the plots, it is clear that \texttt{candies} helps improve the classification done by FETCH, particularly for S/N < 10.}
    \label{fig:your_vs_candies_class}
\end{figure*}

\begin{figure}
    \centering
    \includegraphics[width=0.45\textwidth]{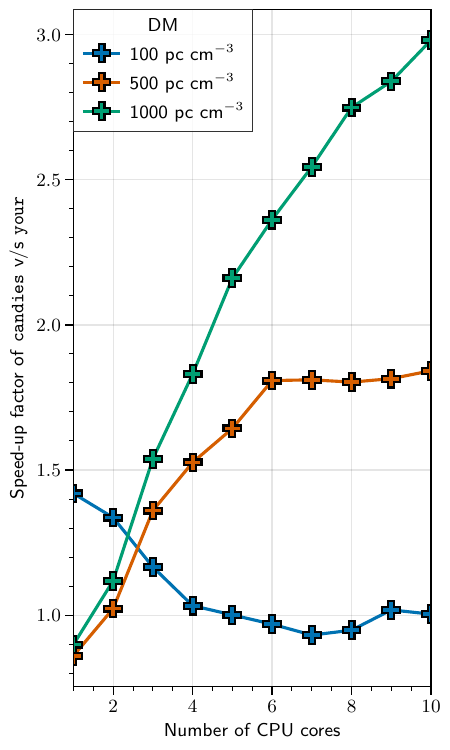}
    \caption{The real-time speed-up factor of \texttt{candies} v/s \texttt{your}, plotted v/s the number of CPU cores used. Note that both \texttt{your} and \texttt{candies} use multiple CPU cores to process several candidates in parallel on the GPU; each candidate is still processed on the GPU. This benchmark was obtained on one of SPOTLIGHT's compute nodes, using one of its NVIDIA A100 GPUs, for 50 candidates in each run. For each point, we take the mean of 3 runs. From the plot, it is clear that \texttt{candies} is 1 to 3 times faster than \texttt{your}.}
    \label{fig:your_vs_candies_speed}
\end{figure}

These features are then passed to \texttt{FETCH}\footnote{\url{https://github.com/devanshkv/fetch}} \citep{agarwal_fetch_2020}, a classifier based on convolutional neural networks (CNNs). \texttt{FETCH} is trained on actual and simulated pulsar and FRB data, and looks for particular characteristics in both features: a straight line spanning the entire band in the dedispersed dynamic spectrum, and a bowtie-like structure in the DM transform. Previously, when \texttt{your} was used to generate the same features for GMRT data, this structure was almost indistinguishable, both by eye, and by \texttt{FETCH} itself. An example of the same can be seen in Fig. \ref{fig:your_vs_candies_zoom}, which shows these features obtained from data taken with Band 3 (300 - 500 MHz), Band 4 (550 - 750 MHz), and Band 5 (1060 - 1460 MHz) of the GMRT, using both \texttt{your} and \texttt{candies}. In Fig. \ref{fig:your_vs_candies_class}, an even clearer comparison can be seen between the two packages; here, we compare how \texttt{FETCH} responds to features generated by either package, for actual single pulses detected from FRB~20180916B and FRB~20201124A using the GMRT. It is evident from Fig. \ref{fig:your_vs_candies_class} that \texttt{candies} helps improve \texttt{FETCH}'s accuracy by a significant margin. It is also faster than \texttt{your}, since it uses custom, optimised GPU kernels to create the required features. A speed comparison between \texttt{your} and \texttt{candies}, as benchmarked on SPOTLIGHT's NVIDIA A100 GPUs, can be seen in Fig. \ref{fig:your_vs_candies_speed}. From the plot, it is clear that \texttt{candies} is $1\times$ to $3\times$ faster than \texttt{your}, and this speed-up with respect to \texttt{your} is higher for larger dispersion measures.

For classification, \texttt{FETCH} uses two CNNs, each trained separately on one of the two features. \texttt{spltpipe} uses \texttt{FETCH}'s best performing model, model \textit{a}, comprised of \texttt{DenseNet121} and \texttt{Xception}. Then, a multiplicative fusion approach is used to fuse the two networks, yielding a single output: a probability indicating how likely it is that the candidate is an astrophysical event. All candidates with a probability of 50\% and above are passed to the next stage. After classification, the number of candidates can range from 1 - 10 per beam per block. \texttt{candies} handles both the feature extraction and classification stages, since it has absorbed the \texttt{ONNX}-based inference-only version of \texttt{FETCH}\footnote{Available here: \url{https://github.com/devanshkv/fetch/tree/onnx}.} entirely. It can work on both SIGPROC filterbank files recorded on disk, as well as on data obtained directly from SPOTLIGHT's ring buffers.

\subsubsection{Triggering}\label{sec:triggering}

All of the above stages are carried out independently on each beam and each node, and the candidate lists from all beams across all nodes are merged into one, and passed through a final candidate optimisation stage. Candidates are grouped across the DM-time plane, and a final list of triggers is generated. Currently, 1 trigger is generated every 10 minutes on average, and thus we expect to process $\sim 100$ such triggers per day. For each trigger, high-resolution visibility data (sampled every 1.31072 milliseconds) and baseband data (sampled at the Nyquist rate) is dumped from ring buffers analogous to the ones for beamformed data, sliced around the trigger's arrival time. The former is immediately processed by SPOTLIGHT's real-time imaging pipeline (Pal et al. (in prep)), while the latter is stored, to be processed at a later stage by our offline correlator (Reddy et al. (in prep)).

\subsection{\texttt{aa.single} and \texttt{splt.single}}\label{sec:iabeam}

The IA beam is searched using single-beam analogues of the same real-time transient search system described above: \texttt{aa.single}\footnote{\url{https://github.com/nsmspotlight/aa.single}} and \texttt{splt.single}\footnote{\url{https://github.com/nsmspotlight/splt.single}}. Most modules remain unchanged, since they work on a per-beam basis; however, some modules are removed, such as the ones for coincidence or anti-coincidence filtering, since they only make sense in the context of multiple beams.

\subsection{\texttt{arachne}}\label{sec:arachne}

\texttt{arachne} is SPOTLIGHT's real-time injection module. It can inject one or multiple events from a library of pre-simulated bursts directly into \texttt{FRBRing}. These events are injected before the data from the ring buffer is processed by \texttt{Ares}. The library of pre-simulated bursts is constructed using \texttt{simulatesearch} \citep{luo_simulating_2022}, but can be constructed using other simulation programs as well. The injection methodology follows that of \cite{luo_simulating_2022}. Assuming Gaussian radiometer noise and using the known frequency-dependent system temperature and telescope gain of the observation, the probability that an injected signal modifies a given data bit is computed and used to generate realistic signal injections. This probability will change based on which quantised level the bit currently occupies, as well as the amplitude of the injected signal, and will need to be determined for each quantised level, including the one the bit already occupies (that is, the probability that the bit will remain unchanged). As \cite{luo_simulating_2022} already states, this calculation is already quite challenging for the case of 2-bit data, since the number of probabilities that need to be estimated is equal to the number of quantised levels at or above the one occupied by each data bit, which can vary from $2^{0} = 1$ to $2^{2} = 4$ for 2-bit data (depending on whether the highest or lowest level is occupied). For the general case, that is, for $\mathcal{M}$-bit data, this can vary from 1 to $2^{\mathcal{M}}$. Since the data obtained from SPOTLIGHT is 8-bit quantised by default, we extend this logic, to both the general, $\mathcal{M}$-bit case, and to the specific 1-bit, 2-bit, and 8-bit cases; the details of this are given in Appendix \ref{appendix:B}.

\section{Performance}\label{sec:performance}

\begin{figure}
    \centering
    \includegraphics[width=0.45\textwidth]{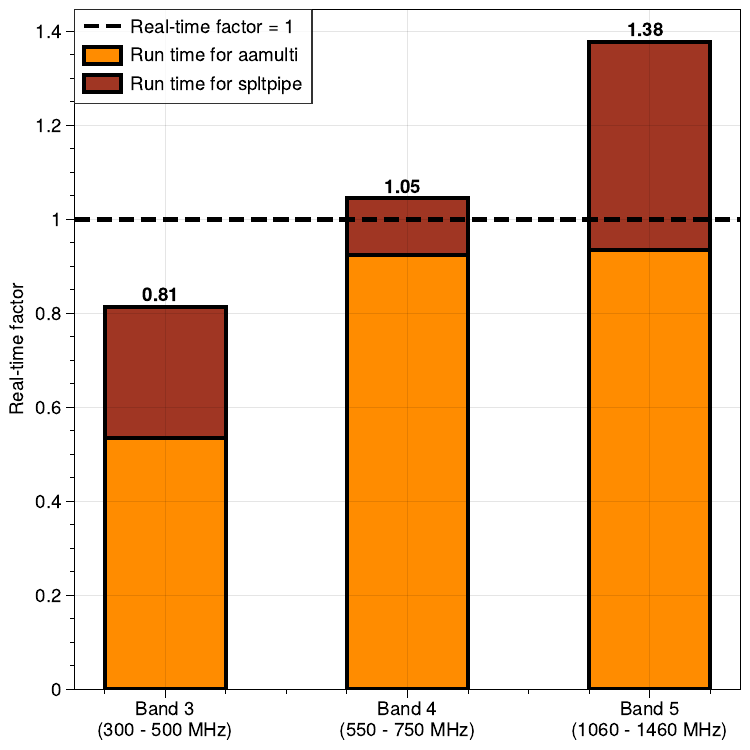}
    \caption{Real-time factors (that is, the time taken by the pipeline versus the time processed by the pipeline) for \texttt{aamulti}, \texttt{spltpipe}, and both combined for GMRT's Band 3 (300 $-$ 500 MHz), Band 4 550 $-$ 750 MHz), and Band 5 (1060 $-$ 1460 MHz). The dashed black line represents a real-time factor of 1; ideally, the pipeline's real-time factor should lie firmly below this limit.}
    \label{fig:realtimefactors}
\end{figure}

Both \texttt{aamulti} and \texttt{spltpipe} output detailed benchmarks with each run. These were used to estimate the real-time factor of each run of the pipeline. Fig \ref{fig:realtimefactors} shows the mean real-time factor for \texttt{aamulti}, \texttt{spltpipe}, and both combined for GMRT's Band 3 (300 $-$ 500 MHz), Band 4 550 $-$ 750 MHz), and Band 5 (1060 $-$ 1460 MHz). We have only used runs from GMRT's Cycle 50, since real-time RFI mitigation was implemented on both voltage and beamformed data from this cycle (Wani et al. (in prep)). In this cycle, the pipeline is running on 640 beams, up to a DM of 2000 pc cm$^{-3}$, and with an S/N limit of 10$\sigma$, which corresponds to a fluence limit of 0.2 Jy ms. From the plot, it is clearly evident that the pipeline is doing as good as, or better than, real-time for both Band 3 and Band 4, and is a bit worse than real-time for Band 5. This is likely due to the fact that dispersive delays are lower in Band 5, which implies that the pipeline cannot easily distinguish between astrophysical signals and RFI, leading to a much larger number of spurious candidates.
\section{Results}\label{sec:results}

The SPOTLIGHT multibeam, real-time transient search pipeline was first provisionally deployed in GTAC Cycle 40 (April 2025), and then deployed after proper science and functionality validation in GTAC Cycle 49 (October 2025) with 160 beams. A major milestone was achieved in GTAC Cycle 50 (April 2026) when the entire SPOTLIGHT system, including the real-time transient search pipeline, was deployed with 640 beams. This deployment also included the incoherent beam real-time transient search pipeline. As of June 2026, the pipeline has detected 2870 bursts from 42 known sources. A census of these detections can be seen in Fig. \ref{fig:census}, where we plot the fluence of each burst versus its dispersion measure (DM). The fluence is estimated using the \texttt{gmrtetc}\footnote{\url{https://github.com/astrogewgaw/gmrtetc}} package, which is a pure Python implementation of GMRT's Exposure Time Calculator. \texttt{gmrtetc} gives an estimate of the RMS for each burst detected in a particular epoch, based on observational parameters, and the width of the burst (as obtained from the search)\footnote{For a more detailed overview of how GMRT's Exposure Time Calculator functions, refer to the help document available here: \url{http://www.ncra.tifr.res.in:8081/\~secr-ops/etc/etc_help.pdf}.}. These estimates are then scaled by the detection S/N to obtain the flux density. The fluence of each burst is then simply a product of the flux density and the width of the burst. Note that the S/N might not take into account all components of a bursts; only the component or components successfully matched by one of the trial boxcars are accounted for. The red dashed line in Fig. \ref{fig:census} represents the theoretical fluence limit of 0.2 Jy ms for the SPOTLIGHT system, estimated for a burst with an S/N $= 10$; it can be clearly seen that many detections reach this limit, proving that the system and the pipeline are working as expected.

Fig. \ref{fig:galactic_map} shows the sky coverage of the SPOTLIGHT system, from the deployment of the real-time pipeline, till June 2026. The pipeline ran for 22,043 scans, covered a total of 277.77 sq. degrees of sky in all bands; 239.09 sq. degrees were covered in Band 3 (300 to 500 MHz), 48.49 sq. degrees were covered in Band 4 (550 to 950 MHz), and 10.51 sq. degrees were covered in Band 5 (1060 to 1460 MHz). The 42 known sources detected by the pipeline so far are overlaid on this sky coverage, and each point's colour represents the DM of each source. Using the all-sky rate of $[523 \pm 30\text{(stat.)}^{+143}_{-131}\text{(sys.)}]$ sky$^{-1}$ day$^{-1}$ estimated from CHIME/FRB's First Catalog \citep{amiri_first_2021, collaboration_erratum_2023}, one can estimate that the pipeline should have detected $2.78^{+0.62}_{-0.36}$ FRBs, assuming a fluence index of $\alpha = -1.4$.
\section{Summary}\label{sec:summary}

We have presented the design, implementation, and on-sky performance of the SPOTLIGHT real-time transient search pipeline, a GPU-accelerated system developed to enable commensal searches for fast radio transients with the uGMRT. Operating on a dedicated high-performance computing facility comprising 90 NVIDIA A100 GPUs and 60 compute servers, the pipeline is deployed to process up to 640 post-correlation beams in real time while simultaneously supporting transient localisation through high-time-resolution visibility and baseband recording.

The pipeline combines brute-force GPU-based dedispersion, matched-filter-based single-pulse searching, candidate clustering, coincidence and anti-coincidence filtering, machine-learning-based classification, and automated triggering. To ensure continuous validation of system performance, we have also developed \texttt{arachne}, a real-time signal injection framework capable of injecting realistic synthetic bursts directly into the live beamformed data stream.

During its initial science deployment in uGMRT Cycles 49 and 50, the system operated commensally with regular telescope observations and successfully detected 2870 bursts from 42 known sources. The detected bursts span a wide range of dispersion measures and fluences, reaching the predicted survey sensitivity limit of approximately 0.2 Jy ms, thereby validating both the sensitivity and robustness of the pipeline. Performance measurements demonstrate near real-time operation across the major uGMRT observing bands, while processing 640 beams and searching dispersion measures up to 2000 pc cm$^{-3}$. The SPOTLIGHT real-time transient search pipeline represents one of the largest GPU-powered commensal transient-search systems currently operating at low radio frequencies.
\begin{acknowledgments}

We acknowledge the invaluable contribution of colleagues from the Oxford e-Research Centre (OeRC), NVIDIA, and the Centre for the Development of Advanced Computing (C-DAC). We thank our colleagues at the Centre for Development of Advanced Computing (C-DAC) for their support in setting up the Param Brahmand data centre at the GMRT. We acknowledge the funding of the SPOTLIGHT backend (called Param Brahmand) under the National Supercomputing Mission's (NSM) Phase 3, as well as the support of the Department of Atomic Energy (DAE), Government of India, under project no. $\rm 12-R\&D-TFR5.02-0700$, for the contributions towards the overhead cost. The uGMRT is operated by the National Centre for Radio Astrophysics of the Tata Institute of Fundamental Research, India.

We gratefully acknowledge support from the ``Building Indo–UK Collaborations Towards the Square Kilometre Array'', funded under the DAE–STFC Technology and Skills Programme, which facilitated the development of SPOTLIGHT's highly efficient, real-time transient search pipelines. We sincerely thank the uGMRT engineers involved in the SPOTLIGHT survey for their relentless efforts in commissioning and stabilising SPOTLIGHT's correlator and beamformer systems, ensuring high quality data flow for all of SPOTLIGHT's real-time and offline systems. We also thank the uGMRT operators for their coordinated efforts in successfully conducting the SPOTLIGHT survey observations. This research was supported in part by the International Centre for Theoretical Sciences (ICTS) for the FTSky: A program in the field of Fast Radio Transients (code: ICTS/FTSky2025/10).

\end{acknowledgments}
\appendix

\section{Derivation for the change in DM for a given S/N loss}\label{appendix:A}

In order to derive Eq. \ref{eq:deltadm}, we need to first calculate the S/N loss due to dispersing a pulse at an incorrect trial DM value. This is a direct consequence of the dispersive smearing discussed in \S\ref{sec:dedispersion}. As per \cite{cordes_searches_2003}, if the DM of a candidate is off by $\delta\mathrm{DM}$, then the ratio of the measured peak flux $S(\delta\mathrm{DM})$ to the true peak flux $S$, for a Gaussian pulse with a full width at half maximum (FWHM) of $W$ is:

\begin{equation}\label{eq:snrloss}
    \frac{S(\delta\mathrm{DM})}{S}
    = \frac{\sqrt{\pi}}{2} \zeta^{-1} \mathrm{erf}(\zeta),
\end{equation}

where $\mathrm{erf}(x)$ is the error function, defined as:

\begin{equation}\label{eq:erf}
    \mathrm{erf}(x) := \frac {2}{\sqrt{\pi}} \int_{0}^{x} e^{-t^{2}}dt
\end{equation}

and $\zeta$ is defined as:

\begin{equation}\label{eq:zetadef}
    \zeta :=
    6.91 \times 10^{-3}
    \delta\mathrm{DM}
    \frac{\Delta\nu_{\mathrm{MHz}}}{W_{\mathrm{ms} \, \nu_{\mathrm{GHz}}^{3}}},
\end{equation}

where $\Delta\nu_{\mathrm{MHz}}$ is the observation bandwidth in MHz, $\nu_{\mathrm{GHz}}$ is the central observational frequency in GHz, and $W_{\mathrm{ms}}$ is the width of the candidate in milliseconds. Note that the error function, $\mathrm{erf}(x)$, approaches 1 as $x \rightarrow \infty$; in fact, this approximation is valid for most values of $x \geq 1$. Thus, we can approximate Eq. \ref{eq:snrloss} as:

\begin{equation}\label{eq:snrlossapprox}
    \frac{S(\delta\mathrm{DM})}{S}
    \approx \frac{\sqrt{\pi}}{2 \zeta}, \text{ if } \zeta \gg 1.
\end{equation}

And thus, substituing the value of $\zeta$ from Eq. \ref{eq:zetadef} into Eq. \ref{eq:snrlossapprox}, we get:

\begin{align*}
    \delta \mathrm{DM}
    &\approx \frac{\sqrt{\pi}}{2 \times 6.91 \times 10^{-3}}
            W_{\mathrm{ms}}
            \left(\frac{S}{S(\delta DM)}\right)
            \frac{\nu_{\mathrm{GHz}}^{3}}{\Delta\nu_{\mathrm{MHz}}} \\
    &= \frac{\sqrt{\pi}}{2 \times 691}
            W_{\mathrm{s}}
            \left(\frac{S}{S(\delta DM)}\right)
            \frac{\nu_{\mathrm{MHz}}^{3}}{\Delta\nu_{\mathrm{MHz}}} \\
    &= \frac{\sqrt{\pi}}{1382}
            W_{\mathrm{s}}
            \left(\frac{S}{S(\delta DM)}\right)
            \frac{\nu_{\mathrm{MHz}}^{3}}{\Delta\nu_{\mathrm{MHz}}},
\end{align*}

which is precisely Eq. \ref{eq:deltadm}.

\section{Injecting a signal into a $\mathcal{M}$-bit quantised data stream}\label{appendix:B}

Assuming that the background noise is represented by a Gaussian distribution, the probability density function (pdf) of the noise $\mathcal{N}$ is:

\begin{equation*}
P(\mathcal{N}) = \frac{1}{\sigma \sqrt{2 \pi}} e^{-\frac{(\mathcal{N} - \mu)^2}{2 \sigma^2}}
\end{equation*}

where the noise level $\sigma$ is given by the radiometer equation for a single antenna:

\begin{equation*}
\sigma = \frac{T_{sys}}{G\sqrt{N_{\text{pol}} \delta t \Delta \nu}}
\end{equation*}

The cumulative distribution function of noise is given by:

\begin{equation*}
\phi(\mathcal{N}) = 
    \frac{1}{2} 
    \left[
        1 + \text{erf}
        \left(
            \frac{\mathcal{N} - \mu}{\sigma \sqrt{2}}
        \right)
    \right]
\end{equation*}

For a given signal $\mathcal{S}$ injected to data samples, we get digitized values ranging from 0 to $2^{\mathcal{M}} - 1$ for the $\mathcal{M}$ bit case. The probability matrix for digit changes upon injecting a signal would be:

\begin{equation*}
P_{ij} =
    \begin{pmatrix}
        P_{00} & P_{01} & \cdots & P_{0(2^{\mathcal{M}} - 1)}\\
        0 & P_{11} & \cdots & P_{1(2^{\mathcal{M}} - 1)}\\
        : &  & \ddots & \\
        0 & 0 & \cdots & P_{(2^{\mathcal{M}} - 2)(2^{\mathcal{M}} - 1)}\\
        0 & 0 & \cdots & 1
    \end{pmatrix}
\end{equation*}

where $P_{nm}$ is the probability to change the bit from state $n$ to the state $m$ and is given by:

\begin{equation*}
P_{nm} = 
    \frac{\text{Probability to go from state $n$ to state $m$}}
         {\text{Probability to remain in original state $n$}}
\end{equation*}

\begin{figure}
    \centering
    \includegraphics[width=0.65\textwidth]{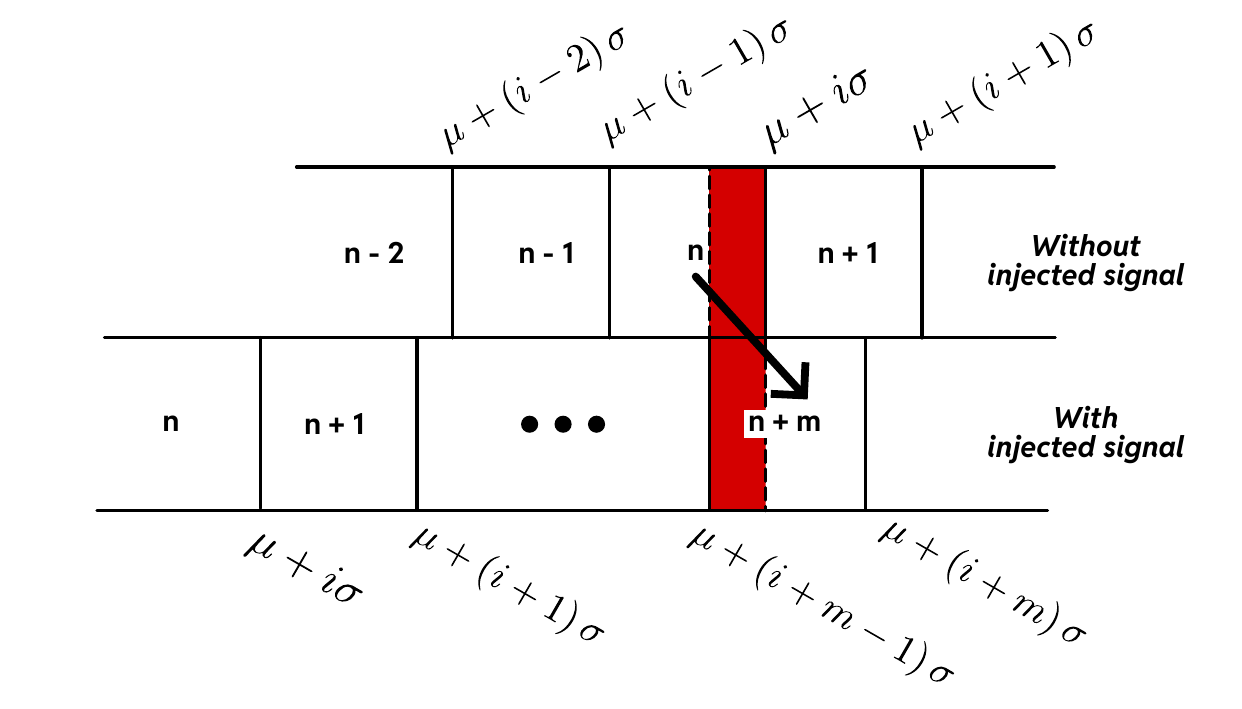}
    \caption{A diagram illustrating how a bit is shifted based on the amount of signal injected. Assuming the bit occupies the $n$th level, the probability of it switching to the $(n + m)$th level is calculated. This probability depends on the amount of overlap between the two levels (marked in red), which, in turn, depends on the amount of signal injected. The bit is then shifted to the level for which this probability is the highest. The same logic is repeated for each bit in the quantised data stream.}
    \label{fig:bitshifts}
\end{figure}

Say we have $\mathcal{M}$ bit data; thus there are $2^{\mathcal{M}}$ levels. Consider state $n$ lying between $\mu + il\sigma$ and $\mu+(i+1)l\sigma$, where $\mu$ is the mean of Gaussian noise, and $l$ is the level setting number for digitization. Here, $i$ runs from  $-(\frac{2^{\mathcal{M}}}{2} - 1)$ to $+(\frac{2^{\mathcal{M}}}{2} - 1)$ and $n$ varies from 0 to $2^{\mathcal{M}} - 1$. As shown in the figure (\ref{fig:bitshifts}), the probability of bit shift can be calculated as,

\begin{equation*}
P_{n \rightarrow n+m}
    = \frac{\text{Area under pdf between intersection of states $n$ and $n+m$}}{\text{Area under pdf up to state $n$}}\\
\end{equation*}

\begin{equation} \label{Pn->n+m}
    P_{n \rightarrow n+m} =
    \frac{
        \phi(\text{min}(\mu + (i+m)l\sigma - \mathcal{S}, \mu + il\sigma))
        -  \phi(\text{max}(\mu + (i+m-1)l\sigma - \mathcal{S}, \mu + (i-1)l\sigma))
    }
    {
        \phi( \mu + il\sigma)
        - \phi( \mu + (i-1)l\sigma)
    }
\end{equation}

where $n$ and $i$ are related as $n - i =\frac{2^{\mathcal{M}}}{2} - 1$. \\

The first state $(0)$ extends from $-\infty$ to $\mu - (\frac{2^{\mathcal{M}}}{2} - 1)l\sigma$ and the last state $(2^{\mathcal{M}} - 1)$ extends from $\mu + (\frac{2^{\mathcal{M}}}{2} - 1)l\sigma$ to $+\infty$. Therefore, we need to accommodate these variation at the boundary states. This can be done by modifying above formula as,

\begin{equation} \label{P0->0}
    P_{0 \rightarrow 0}  
        = \frac{\phi(\mu + (1-\frac{2^{\mathcal{M}}}{2})l\sigma - \mathcal{S})}{\phi( \mu +(1-\frac{2^{\mathcal{M}}}{2})l\sigma)}
\end{equation}

\begin{equation} \label{P0->m}
    P_{0 \rightarrow m} =
        \frac{
            \phi(\text{min}(\mu + (m+1-\frac{2^{\mathcal{M}}}{2})l\sigma - \mathcal{S}, \mu +(1-\frac{2^{\mathcal{M}}}{2})l\sigma))
            -  \phi(\mu + (m-\frac{2^{\mathcal{M}}}{2})l\sigma - \mathcal{S})
        }
        {
            \phi( \mu +(1-\frac{2^{\mathcal{M}}}{2})l\sigma)
        }
\end{equation}

\begin{equation} \label{Pn->max}
    P_{n \rightarrow 2^{\mathcal{M}} - 1} =
        \frac{
            \phi(\mu + (n+1-\frac{2^{\mathcal{M}}}{2})l\sigma)
            -  \phi(\text{max}(\mu + (\frac{2^{\mathcal{M}}}{2} - 1)l\sigma - \mathcal{S}, \mu + (n-\frac{2^{\mathcal{M}}}{2})l\sigma))
        }
        {
            \phi( \mu + (n+1-\frac{2^{\mathcal{M}}}{2})l\sigma)
            - \phi( \mu + (n-\frac{2^{\mathcal{M}}}{2})l\sigma)
        }
\end{equation}
 
\begin{equation} \label{P0->max}
    P_{0 \rightarrow 2^{\mathcal{M}} - 1} =
        \frac{
            \phi( \mu +(1 -\frac{2^{\mathcal{M}}}{2})l\sigma)
            -  \phi(\mu + (\frac{2^{\mathcal{M}}}{2} - 1)l\sigma - \mathcal{S})
        }
        {
            \phi( \mu +(1-\frac{2^{\mathcal{M}}}{2})l\sigma)
        } 
        = 1 - \frac{\phi(\mu + (\frac{2^{\mathcal{M}}}{2} - 1)l\sigma - \mathcal{S})}{\phi( \mu +(1-\frac{2^{\mathcal{M}}}{2})l\sigma)}
\end{equation}

and,

\begin{equation} \label{Pmax->max}
    P_{2^{\mathcal{M}} - 1 \rightarrow 2^{\mathcal{M}} - 1}  
        = 1 
\end{equation}

Equation (\ref{Pn->n+m}) to (\ref{Pmax->max}) are general formulas to calculate probability of bit change from one state to other when a signal $\mathcal{S}$ is injected into the quantized data stream.

Now, \textbf{for the 8-bit case}, we have $n - i = \frac{2^8}{2} - 1 =127$. Thus,

\begin{equation*}    
    P_{n \rightarrow n+m} =
        \frac{
            \phi(\text{min}(\mu + (n+m-127)l\sigma - \mathcal{S}, \mu + (n-127)l\sigma))
            -  \phi(\text{max}(\mu + (n+m-128)l\sigma - \mathcal{S}, \mu + (n-128)l\sigma))
        }
        {
            \phi( \mu + (n-127)l\sigma)
            - \phi( \mu + (n-128)l\sigma)
        }
\end{equation*}
where the level-setting parameter $l$ for 8-bit data is 0.030765, as adopted from \cite{kouwenhoven_effect_2001}. 

The formula is appropriate, given that we are away from the boundary states (the first and last states). Probability of changing bit from $0$ to $0$ and $255$ to $255$, is

\begin{equation}
    P_{0 \rightarrow 0}  
        = \frac{\phi(\mu -127l\sigma - \mathcal{S})}{\phi( \mu -127l\sigma)}
        \text{  and  }
    P_{255 \rightarrow 255} = 1 \text{, respectively. }    
\end{equation}

For bit change from state $0$ to any higher state $m$, the formula modifies as

\begin{equation*}
    P_{0 \rightarrow m} =
        \frac{
            \phi(\text{min}(\mu + (m-127)l\sigma - \mathcal{S}, \mu -127l\sigma))
            -  \phi(\mu + (m-128)l\sigma - \mathcal{S})
        }
        {
            \phi( \mu -127l\sigma)
        }
\end{equation*}
Similarly, for bit change from any non-zero state $n$ to the highest state $255$, the probability is

\begin{equation*}
    P_{n \rightarrow 255} =
        \frac{
            \phi(\mu + (n-127)l\sigma)
            -  \phi(\text{max}(\mu + 127l\sigma - \mathcal{S}, \mu + (n-128)l\sigma))
        }
        {
            \phi( \mu + (n-127)l\sigma)
            - \phi( \mu + (n-128)l\sigma)
        }
\end{equation*}
Lastly, the probability for bit change from the lowest state $0$ to the highest state $255$ is

\begin{equation*}
    P_{0 \rightarrow 255} =
        \frac{
            \phi( \mu -127l\sigma)
            -  \phi(\mu + 127l\sigma - \mathcal{S})
        }
        {
            \phi( \mu -127l\sigma)
        } 
        = 1 - \frac{\phi(\mu + 127l\sigma - \mathcal{S})}{\phi(\mu -127l\sigma)}
\end{equation*}

\textbf{For a 1-bit case}, $\mathcal{M}$ is 1 and thus there are only 2 levels 0 and 1. So, $n$ and $i$ are equal for 1-bit data. Only possible transition between these two states is 0 to 1, and the resulting probability from equation \ref{P0->max} is

\begin{equation} \label{P0to1}
    P_{0 \rightarrow 1} =
        \frac{
            \phi( \mu +(1-\frac{2^{1}}{2})l\sigma)
            -  \phi(\mu + (\frac{2^{1}}{2} - 1)l\sigma - \mathcal{S})
        }
        {
            \phi( \mu +(1-\frac{2^{1}}{2})l\sigma)
        } 
        = 1 - \frac{\phi(\mu - \mathcal{S})}{\phi(\mu)}
        = 1 - \frac{\phi(\mu - \mathcal{S})}{0.5}
\end{equation}
 
\begin{equation} \label{P0to1_2}
    P_{0 \rightarrow 0} = 1 - P_{0 \rightarrow 1}
        = \frac{\phi(\mu - \mathcal{S})}{0.5}
\end{equation}

\begin{equation} \label{P0to1_2}
    P_{1 \rightarrow 0} = 0 ; P_{1 \rightarrow 1} = 1
\end{equation}

\textbf{For a 2-bit case}, $\mathcal{M}$ is 2 and thus there are 4 levels namely 0, 1, 2, and 3. In this case, $n$ and $i$ are related as $n - i = \frac{2^2}{2} - 1=1$. 

Using equation (\ref{Pn->n+m}),

\begin{align} \label{P11}
        P_{1 \rightarrow 1} 
        &=
        \frac{
            \phi(\text{min}(\mu + (i+1)l\sigma - \mathcal{S}, \mu + il\sigma))
            -  \phi(\text{max}(\mu + (i+1-1)l\sigma - \mathcal{S}, \mu + (i-1)l\sigma))
        }
        {
            \phi( \mu + il\sigma)
            - \phi( \mu + (i-1)l\sigma)
        }\\ 
        &=
        \frac{
            \phi(\mu  - \mathcal{S})
            -  \phi( \mu -l\sigma)
        }
        {
            \phi( \mu )
            - \phi( \mu -l\sigma)
        }\\
        &=
        \frac{
            \phi(\mu  - \mathcal{S})
            -  \phi( \mu -l\sigma)
        }
        {
            0.5
            - \phi( \mu -l\sigma)
        }
\end{align}

\begin{align} \label{P12}
        P_{1 \rightarrow 2} 
        &=
        \frac{
            \phi(\text{min}(\mu + (i+1)l\sigma - \mathcal{S}, \mu + il\sigma))
            -  \phi(\text{max}(\mu + (i+1-1)l\sigma - \mathcal{S}, \mu + (i-1)l\sigma))
        }
        {
            \phi( \mu + il\sigma)
            - \phi( \mu + (i-1)l\sigma)
        }\\ 
        &=
        \frac{
            \phi(\text{min}(\mu + l\sigma - \mathcal{S}, \mu))
            -  \phi(\text{max}(\mu - \mathcal{S}, \mu -l\sigma))
        }
        {
            \phi( \mu )
            - \phi( \mu -l\sigma)
        }\\
        &=
        \frac{
            \phi(\text{min}(\mu + l\sigma - \mathcal{S}, \mu))
            -  \phi(\text{max}(\mu - \mathcal{S}, \mu -l\sigma))
        }
        {
            0.5
            - \phi( \mu -l\sigma)
        }
\end{align}

and,

\begin{align} \label{P22}
        P_{2 \rightarrow 2} 
        &=
        \frac{
            \phi(\text{min}(\mu + (i+1)l\sigma - \mathcal{S}, \mu + il\sigma))
            -  \phi(\text{max}(\mu + (i+1-1)l\sigma - \mathcal{S}, \mu + (i-1)l\sigma))
        }
        {
            \phi( \mu + il\sigma)
            - \phi( \mu + (i-1)l\sigma)
        }\\ 
        &=
        \frac{
            \phi(\mu + l\sigma - \mathcal{S})
            -  \phi(\mu)
        }
        {
           \phi( \mu +l\sigma)
           - \phi(\mu ) 
        }\\
        &=
        \frac{
            \phi(\mu + l\sigma - \mathcal{S})
            -  0.5
        }
        {
           \phi( \mu +l\sigma)
           - 0.5 
        }
\end{align}

Using equation (\ref{P0->0}),

\begin{align} \label{P00}
        P_{0 \rightarrow 0}  
        = \frac{\phi(\mu + (1-\frac{2^{\mathcal{M}}}{2})l\sigma - \mathcal{S})}{\phi( \mu +(1-\frac{2^{\mathcal{M}}}{2})l\sigma)}
        = \frac{\phi(\mu -l\sigma - \mathcal{S})}{\phi( \mu -l\sigma)}
\end{align}

From equation (\ref{P0->m}),

\begin{equation*}
    P_{0 \rightarrow m} =
        \frac{
            \phi(\text{min}(\mu + (m-1)l\sigma - \mathcal{S}, \mu -1l\sigma))
            -  \phi(\mu + (m-2)l\sigma - \mathcal{S})
        }
        {
            \phi( \mu -1l\sigma)
        }
\end{equation*}

\begin{equation} \label{P01}
    P_{0 \rightarrow 1} =
        \frac{
            \phi(\text{min}(\mu - \mathcal{S}, \mu -l\sigma))
            -  \phi(\mu - l\sigma - \mathcal{S})
        }
        {
            \phi( \mu -l\sigma)
        }
\end{equation}

\begin{equation} \label{P02}
    P_{0 \rightarrow 2} =
        \frac{
            \phi(\text{min}(\mu +l\sigma - \mathcal{S}, \mu -l\sigma))
            -  \phi(\mu - \mathcal{S})
        }
        {
            \phi( \mu -l\sigma)
        }
\end{equation}

from equation (\ref{Pn->max}).

\begin{equation*}  
    P_{n \rightarrow 3} =
        \frac{
            \phi(\mu + (n-1)l\sigma)
            -  \phi(\text{max}(\mu +l\sigma - \mathcal{S}, \mu + (n-2)l\sigma))
        }
        {
            \phi( \mu + (n-1)l\sigma)
            - \phi( \mu + (n-2)l\sigma)
        }
\end{equation*}

\begin{equation} \label{P13}
    P_{1 \rightarrow 3} =
        \frac{
            \phi(\mu)
            -  \phi(\text{max}(\mu +l\sigma - \mathcal{S}, \mu -l\sigma))
        }
        {
            \phi( \mu)
            - \phi( \mu -l\sigma)
        }=
        \frac{
            0.5
            -  \phi(\text{max}(\mu +l\sigma - \mathcal{S}, \mu -l\sigma))
        }
        {
            0.5
            - \phi( \mu -l\sigma)
        }
\end{equation}

\begin{equation} \label{P23}
    P_{2 \rightarrow 3} =
        \frac{
            \phi(\mu + l\sigma)
            -  \phi(\text{max}(\mu +l\sigma - \mathcal{S}, \mu))
        }
        {
            \phi( \mu + l\sigma)
            - \phi(\mu)
        }=
        \frac{
            \phi(\mu + l\sigma)
            -  \phi(\text{max}(\mu +l\sigma - \mathcal{S}, \mu))
        }
        {
            \phi( \mu + l\sigma)
            - 0.5
        }
\end{equation}

from equation (\ref{P0->max})

\begin{equation} \label{P03}
    P_{0 \rightarrow 3} =
        \frac{
            \phi( \mu -l\sigma)
            -  \phi(\mu + l\sigma - \mathcal{S})
        }
        {
            \phi( \mu -l\sigma)
        } 
\end{equation}

and lastly from equation (\ref{Pmax->max})

\begin{equation} \label{P33}
    P_{3 \rightarrow 3} = 1 
\end{equation}

Equations (\ref{P0to1}) to (\ref{P33}) match with the probability formulas for injection in 1-bit and 2-bit data given in \cite{luo_simulating_2022}, Appendix B. As per the bit shift logic derived above, the formulae for the calculation of the $P_{2 \rightarrow 3}$ and $P_{3 \rightarrow 3}$ probabilities derived in \cite{luo_simulating_2022}, Appendix B, are incorrect, and the corrected formulae are given above in equations \eqref{P23} and \eqref{P33}.

\bibliography{refs}{}
\bibliographystyle{aasjournal}

\end{document}